# High-entropy grain boundaries

Jian Luo[*] and Naixie Zhou

Department of Nanoengineering; Program of Materials Science and Engineering, University of California San Diego, La Jolla, California, U.S.A.

**Abstract**

Do high-entropy alloys and ceramics have their grain boundary (GB) counterparts? As the concept of high-entropy grain boundaries (HEGBs) was initially proposed in 2016, this article provides the first complete and rigorous discussion of the underlying interfacial thermodynamics. A simplified segregation model can illustrate both GB and bulk high-entropy effects, which reduce GB energy with increasing temperature for saturated multicomponent (conventional and high-entropy) alloys. HEGBs can be utilized to stabilize nanocrystalline alloys at high temperatures via thermodynamic and kinetic effects. GB structural disordering and transitions offer further opportunities to attain higher effective GB entropies. Future perspective is discussed.

[*] Email: jluo@alum.mit.edu

High-entropy alloys and ceramics (HEAs and HECs), which are subgroups of broader classes of compositionally complex (or complex concentrated) alloys and ceramics (CCAs and CCCs), have attracted substantial research interests recently.[1-7] Here, we post an interesting scientific question: Can an interface, particularly a grain boundary (GB), be "high entropy"? In 2016, we first proposed the concept and terminology of high-entropy grain boundaries (HEGBs) with a single numerical example in a *Current Opinion* article as a future perspective.[8] We subsequently showed that HEGBs can be utilized to stabilize nanocrystalline alloys (nanoalloys) at high temperatures.[9] Although that original work,[9] particularly the idea of stabilizing nanoalloys with high-entropy effects, has caught substantial interests,[10-20] a rigorous thermodynamic theory of HEGBs has not been presented. This article elaborates the relevant concepts and theory rigorously and present a complete thermodynamic framework for the first time, and subsequently discusses the future perspective.

In a multicomponent system, a GB can certainly be compositionally complex. In fact, even if the bulk phase a conventional multicomponent alloy that is a dilute solid solution of multiple solute components, the GB can be a complex concentrated solution (Fig. 1c), thereby potentially being high entropy even if the bulk alloy is not. Here, we need to first discuss what is GB entropy and thermodynamic characters of HEGBs based on rigorous interfacial thermodynamics.

**GB Entropy and Interfacial Thermodynamics**

There are two general ways to define GB entropy. In the classical Gibbs adsorption theory, we can define the GB excess of entropy ($s^{xs}$), which has a well-defined value (independent of the Gibbs dividing plane). For a bulk phase of a closed system, entropy ($S$) measures how the Gibbs free energy ($G$) decreases with temperature ($T$) at a constant pressure ($P$): $S = -(\partial G/\partial T)_P$. Analogously, we look for a HEGB to have large $s_{GB}^{\text{eff.}} \equiv -(\partial \gamma_{GB}/\partial T)_P$, where GB energy ($\gamma_{GB}$) decreases with increasing temperature. This also represents a beneficial feature to stabilize nanoalloys at high temperatures. However, a multicomponent GB is more complex (than a bulk phase) because it is not a closed system. Different scenarios will be discussed subsequently.

In an *N*-component system, interfacial energy ($\gamma$) is the interfacial excess of grand potential:

$$\gamma = u^{xs} - Ts^{xs} - \sum_{i=1}^{N} \mu_i \Gamma_i, \tag{1}$$

where $u^{xs}$, $s^{xs}$, and $\Gamma_i$ are the interfacial excess of internal energy, entropy, and (adsorption of) the *i*-th component (*i* = 1, 2, …, *N*), respectively, based on the Gibbs definition. For a phase boundary, we can select the Gibbs dividing plane so that $\Gamma_1 = 0$. However, this convention cannot be adopted for a GB, where two abutting grains are the same phase, so that the GB excess quantities are independent of the position of the Gibbs dividing plane. The generalized Gibbs adsorption equation states:

$$d\gamma_{GB} = -s^{xs}dT + \bar{V}dP - \sum_{i=1}^{N} \Gamma_i d\mu_i. \tag{2}$$

For a specific GB in a unary system, its GB energy varies with temperature at a constant pressure:

$$\left(\frac{\partial \gamma_{GB}}{\partial T}\right)_P = -s^{xs} - \Gamma\left(\frac{d\mu}{dT}\right)_P = -\left(s^{xs} - \Gamma\bar{S}\right), \tag{3}$$

where $\bar{S}$ is molar entropy. Since $\Delta v_{\text{GB free}} = -\Gamma \bar{V} = (\partial \gamma_{\text{GB}}/\partial P)_T$ is the GB "free volume" per unit area (in a unit of length), we may define the effective GB entropy for a unary GB as:

$$s_{\text{GB}}^{\text{eff.}} \equiv -\left(\frac{\partial \gamma_{GB}}{\partial T}\right)_P = s^{\text{xs}} + (\bar{S}/\bar{V}) \cdot \Delta v_{\text{GB free}} = s^{\text{xs}} + S_V \cdot \Delta v_{\text{GB free}}, \quad (4)$$

where $S_V$ is the volumetric entropy. It is interesting to note that $s_{\text{GB}}^{\text{eff.}} \equiv -(\partial \gamma_{\text{GB}}/\partial T)_P \neq s^{\text{xs}}$ even for a simple unary GB (because $\Gamma \neq 0$). This inequality is a unique character of unary GBs since we would typically select the Gibbs dividing plane to ensure $\Gamma = 0$ for other types of unary interfaces between two phases, so that $s_{\text{GB}}^{\text{eff.}} = s^{\text{xs}}$ for those cases. For an unary GB, $s_{\text{GB}}^{\text{eff.}}$ is typically positive because the GB region is usually more disordered ($s^{\text{xs}} > 0$) and less dense ($\Gamma < 0$ and $\Delta v_{\text{GB free}} > 0$) than the crystal. This suggests that GB energy generally decreases with increasing temperature or $(\partial \gamma_{\text{GB}}/\partial T)_P < 0$ for a unary GB, as evident by the atomistic simulation of a Ni GB by Foiles (Fig. 1a).[21] If a first-order GB premelting transition occurs, a GB complexion[22] (*a.k.a.* 2D interfacial phase) with a high $s_{\text{GB}}^{\text{eff.}}$ can form (Fig. 1a).

For a multicomponent system, the Gibbs-Duhem equation states:

$$\sum_{i=1}^{N} X_i^{\text{bulk}} d\mu_i = -\bar{S}dT + \bar{V}dP, \quad (5)$$

Combining Eqs. (2) and (5) at a constant pressure ($dP = 0$) produces:

$$d\gamma_{\text{GB}} = -\left[s^{\text{xs}} - \left(\frac{\Gamma_1}{X_1^{\text{bulk}}}\right)\bar{S}\right]dT - \sum_{i=2}^{N}\left[\Gamma_i - \left(\frac{X_i^{\text{bulk}}}{X_1^{\text{bulk}}}\right)\Gamma_1\right]d\mu_i. \quad (6)$$

For an ideal solution, we can obtain:

$$\left(\frac{\partial \gamma_{\text{GB}}}{\partial X_i^{\text{bulk}}}\right)_{P,T,X_j(j\neq 1;j\neq i)} = -\frac{RT}{X_i^{\text{bulk}}}\left[\Gamma_i - \left(\frac{X_j^{\text{bulk}}}{X_1^{\text{bulk}}}\right)\Gamma_1\right] = -RT\left[\frac{\Gamma_i}{X_i^{\text{bulk}}} - \frac{\Gamma_1}{X_1^{\text{bulk}}}\right]. \quad (7)$$

where $R$ is gas constant. Taking the principal element as Component 1, Eq. (7) suggests that increasing amount of segregation ($\Gamma_i > 0$, while $\Gamma_1 < 0$) can reduce GB energy $\gamma_{GB}$. This is illustrated by a binary example of Bi-doped Ni in Fig. 1b, which further shows the occurrence of a first-order GB adsorption transition that accelerates the GB energy reduction with increasing doping.[23]

Let us consider the temperature dependence of GB energy, where we can treat two limiting cases. First, for a fixed bulk composition (*i.e.*, below the solvus or solubility limit for all components), applying the chain rule to Eq. (2) produces:

$$\left(\frac{\partial \gamma_{\text{GB}}}{\partial T}\right)_{P,\mathbf{X}^{\text{bulk}}(X_j^{\text{bulk}},j=1,2,...,N)} = -s^{\text{xs}} - \sum_{i=1}^{N}\Gamma_A\left(\frac{\partial \mu_i}{\partial T}\right)_{P,\mathbf{X}^{\text{bulk}}} = -\left(s^{\text{xs}} - \sum_{i=1}^{N}\Gamma_i \bar{S}_i\right), \quad (8)$$

where $\bar{S}_i$ is the partial molar entropy of the *i*-th component. For a strong segregating system, GB energy often increases with increasing temperature because of temperature-induced desorption that reduces $\gamma_{\text{GB}}$, thereby resulting in a negative effective $s_{\text{GB}}^{\text{eff. (fixed }\mathbf{X}^{\text{bulk}})}$ (while $s^{\text{xs}}$ can still be positive).

Second, we may consider a case of a *saturated* conventional multicomponent alloy with one principal and $(N-1)$ segregating solute components. Here, with changing temperature, the composition of the grains moves along the maximum solvus line, where the chemical potentials of all solutes are pinned by a set of $(N-1)$ precipitates (assumed, for simplicity, to be stoichiometric line compounds). If chemical potentials $\mu_j$ ($j \neq 1$) are fixed (as an approximation), we can derive from Eq. (6):

$$\left(\frac{\partial \gamma_{GB}}{\partial T}\right)_{P, X_j(T) \text{ on solvus}} \approx \left(\frac{\partial \gamma_{GB}}{\partial T}\right)_{P, \mu_j(j \neq 1)} = -\left[s^{xs} - \left(\frac{\Gamma_1}{X_1^{bulk}}\right)\overline{S}\right] \equiv -s_{GB}^{\text{eff. (saturated)}} . \quad (9)$$

For a segregating system, $s_{GB}^{\text{eff. (saturated)}}$ is likely positive since we typically expect: $s^{xs} > 0$ and $\Gamma_1 < 0$.

**GB High-Entropy Effects in a Simplified Segregation Model**

To illustrate GB and bulk high-entropy effects, we can use a statistical thermodynamic model for multicomponent GB segregation (*a.k.a.* adsorption)[8] that is a generalization of the binary Wynblatt-Chatain model[24] with a few simplifications. Considering a general twist GB with segregation limited to the two layers at the GB core and further assuming an ideal solution for simplicity (that can be readily refined to include multilayer adsorption and regular-solution interactions[8]), we can obtain:

$$\begin{cases} \gamma_{GB} = \gamma_{GB}^{(0)} + 2n_{PD}\sum_{i=2}^{N} X_i^{GB}\Delta g_{i \to 1}^{seg.} + 2n\sum_{i=1}^{N} kTX_i^{GB}\ln\left(\frac{X_i^{GB}}{X_i^{bulk}}\right) \\ \frac{X_i^{GB}}{X_i^{bulk}} = \frac{X_1^{GB}}{X_1^{bulk}}\exp\left(-\frac{\Delta g_{i \to 1}^{seg.}}{kT}\right) \end{cases}, \quad (10)$$

where $\gamma_{GB}^{(0)}$ is the GB energy of the pure Component 1, $n_{PD}$ is the planar density of atoms, and $\Delta g_{i \to 1}^{seg.}$ is the free energy (mostly enthalpy) change of GB segregation by swapping an atom of the *i*-th element inside the grain with an atom of the 1$^{st}$ element at the GB. Here, we adopt the sign convention that $\Delta g_{i \to 1}^{seg.} \approx \Delta h_{i \to 1}^{seg.} < 0$ for positive GB segregation (while noting that positive $\Delta h_{i \to 1}^{seg.}$ values are often cited in literature for GB enrichment). We also note that this model does not consider interfacial structural disorder so that $\gamma_{GB}^{(0)}$ is independent of temperature (albeit it should exhibit a negative temperature dependence due to interfacial disordering and GB free volume, as shown in Fig. 1a). Eq. (10) suggests no "high-entropy effect" for $s_{GB}^{\text{eff. (fixed }\mathbf{X}^{bulk})}$ (that is likely negative due to temperature-induced desorption) in this model without considering GB disordering: *e.g.*, in a simple analysis assuming $(N-1)$ segregating solutes of a fixed total amount of solutes $\sum_{i=2}^{N} X_i^{bulk}$ and identical $\Delta h_{i \to 1}^{seg.}$ (and therefore $X_i^{GB}/X_i^{bulk}$), both $\gamma_{GB}$ and $s_{GB}^{\text{eff. (fixed }\mathbf{X}^{bulk})}$ are independent of *N*. However, $s^{xs}$ and $s_{GB}^{\text{eff. (saturated)}}$ can be positive and increase with increasing *N*.

To analytically illustrate a GB high-entropy effect in a dilute multicomponent alloy, we can analyze a hypothetic ideal solution with one principal (M) and $(N-1)$ segregating solute (S$_i$) elements (*i.e.*, the bulk phase is not a HEA, referred to as "Type I" here). For a saturated *N*-component alloy in equilibrium with $(N-1)$ precipitates (*i.e.*, the bulk composition moves along the maximum solvus line following $X_{i,\text{slovus}}^{bulk} \approx \exp(\Delta g_i^{\text{sol./ppt.}}/kT)$ for $i = 2,3,...,N$, where $\Delta g_i^{\text{sol./ppt.}} < 0$ is the free energy of dissolving *i* from the precipitate), we can derive an approximated expression at the dilute limit ($\sum_{i=2}^{N} X_i^{\text{bulk}} \ll 1$) as:

$$\begin{cases} \dfrac{X_i^{\text{GB}}}{X_1^{\text{GB}}} \approx \exp\left(-\dfrac{\Delta g_i^{\text{seg.-ppt.}}}{kT}\right) \\ \gamma_{\text{GB}} - \gamma_{\text{GB}}^{(0)} \approx 2n_{\text{PD}} kT \ln X_1^{\text{GB}} \approx -2n_{\text{PD}} kT \ln\left[1 + \sum_{i=2}^{N} \exp\left(-\dfrac{\Delta g_i^{\text{seg.-ppt.}}}{kT}\right)\right] \end{cases}, \quad (11)$$

where $\Delta g_i^{\text{seg.-ppt.}}$ ($=\Delta g_{i\to 1}^{\text{seg.}} - \Delta g_i^{\text{sol./ppt.}}$) represents the free energy (mostly enthalpy) difference between the segregation and precipitation per atom. Here, an empirical relationship exists: $\Delta g_i^{\text{seg.-ppt.}} \approx -(0.10 \pm 0.06)$ eV/atom (or $-(10 \pm 6)$ kJ/mol).[25] Eq. (11) suggests that adding more segregating elements ($N \uparrow$) can reduce $\gamma_{GB}$ along the solvus line for a $N$-component system (*e.g.*, if $\Delta g_i^{\text{seg.-ppt.}}$ is identical, the sum in Eq. (11) is proportionally to $(N-1)$), representing a GB high-entropy effect.

Furthermore, we can conduct numerical experiments (beyond the dilute limit approximation) to show this GB high-entropy effect using typical parameters for transition metals (the same bonding energy and crystallographic parameters as the case shown in Fig. 6 in Ref. 8, but for an ideal solution in equilibrium with $(N-1)$ precipitates of 1:1 stoichiometric $(M)_1(S_i)_1$ of identical $\Delta g_i^{\text{sol./ppt.}}$ of $-0.13$ eV/atom). Here, we adopt an intermediate $\Delta g_{i\to 1}^{\text{seg.}} = -0.25$ eV/atom (similar to those for Nb or Ta in Ni). Fig. 2a shows this GB high-entropy effect of enhanced reduction in $\gamma_{GB}$ with more segregating elements. Similar results were obtained in a regular-solution model previously.[8]

**Bulk High-Entropy Effects and Type II HEGBs**

In addition, we can show a bulk high-entropy effect for HEAs with multiple principal elements (*i.e.*, the grains are HEAs) and at least one (or more) segregating elements to form "Type II" HEGBs (Fig. 2b).[9] Prior numerical experiments (Fig. 2b)[9] of a hypothetic symmetric ideal solution with $(N-1)$ principal ($M_i$, assumed to have identical properties for simplicity) and one (S) segregating elements based on the same model described above illustrate that $\gamma_{GB}$ can also be reduced via a bulk high-entropy effect. This bulk high-entropy effect lowers the bulk chemical potentials to suppress precipitation to promote more GB adsorption for the HEAs saturated with the segregating element S. Moreover, increasing the number of principal elements in HEAs enhances this bulk high-entropy effect to attain more negative $(\partial \gamma_{GB} / \partial T)_{\text{saturated}}$, as shown in Fig. 2b. Akin to the prior case, we can also derive an approximated analytical expression for an ideal solution at the dilute limit ($X_N^{\text{bulk}} \ll 1$; $X_i^{\text{bulk}} \approx 1/(N-1)$ for $i=1,2,...,(N-1)$) for a Type II HEGB in a HEA:

$$\begin{cases} X_N^{\text{bulk}} = (N-1)^{x/y} \exp\left(\dfrac{\Delta g_N^{\text{sol./ppt.}}}{kT}\right) \\ \gamma_{\text{GB}} - \gamma_{\text{GB}}^{(0)} \approx -2n_{\text{PD}} kT \ln\left[1 + (N-1)^{x/y} \exp\left(-\dfrac{\Delta g_N^{\text{seg.-ppt.}}}{kT}\right)\right] \end{cases}. \quad (12)$$

Here we assume, for simplicity, all binary $i$-$N$ systems ($i=1,2,...,(N-1)$) have identical thermodynamic parameters and the binary solvus lines are pinned by stoichiometric $(M_i)_x(S)_y$ precipitates following $X_{i,\text{binary solvus}}^{\text{bulk}} \approx \exp(\Delta g_N^{\text{sol./ppt.}}/kT)$. Eq. (12) suggests that more principal elements ($N \uparrow$) can also reduce $\gamma_{GB}$ along the solvus line, representing this bulk high-entropy effect for Type II HEGBs with HEA grains.

We should note that this bulk high-entropy effect and Type II HEGBs may also exist in conventional HEAs if one (or more) of the principal element(s) segregates at GBs. An assessment by Miracle and Senkov[3] suggested that slow bulk diffusion in HEAs may not supported by the available data; in fact, the term "sluggish kinetics" was initially introduced based on the observations of fine grains in HEAs. Here, we hypothesize that the "sluggish" grain growth kinetics can exist HEAs because of this bulk high-entropy effect (with the formation of Type II HEGBs).

**Interfacial Structural Disordering and GB Transitions**

The segregation model discussed above do not consider GB structural disordering, which can further increase GB entropy. In addition, first-order GB transitions (*e.g.*, a first-order GB segregation transition for Ni-Bi [26] shown Fig. 1b and a hypothetic first-order GB premelting transition shown in Fig. 1a) can take place to accelerate the reduction in $\gamma_{GB}$. GB structural disordering and segregation can be coupled, as observed in W-Ni (Fig 1c).[27] It is yet unknown when (and why) ordered (Ni-Bi)[26] *vs.* disordered (W-Ni)[27] segregation structures form at GBs. We anticipate that structurally disordered HEGBs can possess even higher effective GB entropies with increased interfacial width (Fig. 1c). This is an area that is largely unexplored and will be discussed further in the Outlook section.

**Stabilizing Nanoalloys at High-Temperatures via HEGBs**

While HEGBs may have several unique characters, one potential application is to utilize them to stabilize nanoalloys at high temperatures.[9] To test this, we modeled, and subsequently fabricated and tested, several Type I (*e.g.*, $Ni_{79.5}Ta_{5.5}Nb_{10}W_5$) and Type II (*e.g.*, $Ni_{29}Fe_{23}Co_{23}Cr_{23}Zr_2$) multicomponent nanoalloys to show substantially improved high-temperature stability (Fig. 2c-e).[9]

Here, we can simulate multicomponent alloys via adopting (1) the above simplified GB segregation model, (2) estimated GB and bonding energies as well as $\Delta g_{i \to 1}^{\text{seg.}}$ for the actual alloys, and (3) binary solvus lines based on empirical $\Delta g_i^{\text{seg.-ppt.}}$ or experimental phase diagrams to forecast useful trends.[9] Here, we use subscript "*" (*e.g.*, "$Ni*_{25}Fe*_{23}Co*_{23}Cr*_{23}Mo*_2Nb*_2Zr*_2$") to denote that they are ideal solutions that mimic the real alloys. Fig. 2c suggests both GB and bulk high-entropy effects in reducing $\gamma_{GB}$, in comparison with benchmarks, for several alloys.[9] Subsequent experiments found that both Type I and Type II nanoalloys with HEGBs exhibit improved (sometime superior) high-temperature stabilities.[9] For example, $Ni_{25}Fe_{23}Co_{23}Cr_{23}Mo_2Nb_2Zr_2$ (with four principal and three segregating elements) maintained ~45 nm grain size after 5 h annealing, outperforming not only their binary counterparts ($Ni_{79}Zr_{21}$ and $Ni_{93}Zr_7$) but also the conventional HEA counterpart ($Ni_{25}Fe_{25}Co_{25}Cr_{25}$), as shown in Fig. 2d and 2e.[9] This modeling and experimental validation methodology can be extended to other alloys in future studies.

We should note that there are two general approaches to inhibit grain growth: (1) thermodynamic stabilization by reducing $\gamma_{GB}$ via solute segregation can reduce the grain growth driving force and (2) kinetic stabilization by solute drag and/or Zener (particle) pinning.[28,29] Notably, in the cases shown in Fig. 2c, GB energies, although being reduced by the HEGBs, are still significant (certainly non-zero). Thus, kinetic stabilization can exist and be important.

A $Ni_{80}Mo_{6.6}Ti_6Nb_6Ta_{1.4}$ alloy was designed to further examine kinetic effects, where the Ti, Nb and Ta contents were about 60-70% of the binary solvus at 900 °C while Mo was ~25% of binary solid solubility.[9] Thus, this case resembles the fixed bulk composition scenario represented by Eq. (8) with no anticipated GB high-entropy effect. However, a prior experiment showed that this nanoalloy possessed a superior high-temperature stability, maintaining 45 nm grain size after annealing 1000 °C for 5h (Fig. 3d and 3e).[9] Here, the kinetic effects, particularly solute drag[30,31] (since binary phase diagrams suggest no precipitate above ~700°C), presumably played the key role in stabilizing this nanoalloy.

Fig. 3 shows that a stable nanoalloy can be achieved via a balance between the thermodynamic driving pressure ($\gamma_{GB}/\langle d \rangle$, where $\langle d \rangle$ is the mean grain size) and critical kinetic solute-drag pressure ($P_C^{drag}$, *i.e.*, the maximum drag force solutes can imposed on the GB before it tears itself off in the Cahn-Lücke-Stüwe model[30,31]): $\gamma_{GB}/\langle d \rangle = P_C^{drag}$. However, both the thermodynamic stabilization and kinetic solute drag become less effective at high temperatures due to temperature-induced GB desorption, which moves the balance point in Fig. 3 towards right. Here, HEGBs may boost GB adsorption to counter the thermally induced desorption (to move the balance point in Fig. 3 towards left to allow a bigger stable window). Specifically, HEGBs can accommodate a higher total amount of GB adsorption within the bulk solid solubility limit (enhanced at high temperatures with more components, as shown in Fig. 2a and 2b). This effect can not only reduce the thermodynamic driving force for grain growth by decreasing $\gamma_{GB}$ at high temperatures (with negative $(\partial \gamma_{GB}/\partial T)_{saturated}$), but also enlarge solute drags to pin grain growth kinetically in "sluggish" HEGBs.

**Outlook**

The theory of HEGBs is still in its infancy stage. Although the concept of HEGBs was proposed in 2016 with some modeling and experimental supports,[8,9] which has caught substantial interests since then,[10-20,32-34] this current article is in fact the first complete and rigorous discussion of the underlying thermodynamic theory. Thermodynamic and kinetic models of HEGBs should be developed and improved to guide this exploration and enable rigorous data analysis.

The multicomponent segregation model presented here, although they can illustrate important concepts and predict some useful trends for transition metal nanoalloys, is highly simplified. Multilayer segregation and regular-solution interactions can be readily included, but analysis of a multicomponent system with a large number of parameters posts a challenge. GB structural disordering, which is not considered in this segregation model, offers a further opportunity to attain even higher effective GB entropies. For example, while this segregation model without considering interfacial disordering suggests no GB high-entropy effect for the fixed grain composition scenario, coupled interfacial disordering and segregation of multiple species, which can enhance each other, may alter the high-temperature behaviors to allow decreasing $\gamma_{GB}$ (and increasing total adsorption amount) with increasing temperature even with a fixed grain composition. This possibility is yet to be tested experimentally. A more rigorous model of HEGBs should consider both chemical and structural (disordering) effects. Developing models for HEGBs in ceramics, which are more prone to form amorphous-like (structurally disordered) GBs,[35] represents another challenge and opportunity.

Here, a future opportunity is to utilize GB disordering and transitions to tailor HEGBs and subsequently the high-temperature stability of nanoalloys (and other properties). Notably, Rupert's group showed the formation of amorphous-like GBs in W-doped Ni,[36] Cu-Zr-Hf,[37] and most recently in Cu-Zr-Hf-Nb-Ti (inset in Fig. 1c) and similar quinary nanoalloys[20] that enhance their high-temperature stability. This suggests a possibility to form structurally disordered (amorphous-like) HEGBs with even higher GB entropies to further enable exceptional high-temperature nanoalloy stability and other exotic properties (*e.g.*, improved mechanical properties with amorphous-like GBs[38,39]). Amorphous-like intergranular films, first discovered in W-Ni (Fig. 1c) and Mo-Ni for metals[27,40] but are more common in ceramics,[35] can accommodate more adsorption. A series of other GB complexions (2D interfacial phases)[22,26,41-44] may also exist in HEGBs, leading to different properties and opportunities.

Notably, Weismuller proposed the existence of an "equilibrium" grain size when the effective $\gamma_{GB}$ approaches zero;[45] yet, Kirchheim[46] suggested (based on the empirical relationship $\Delta G_i^{\text{seg.-ppt.}} \approx -(10 \pm 6)$ kJ/mol [25] that the equilibrium grain sizes in binary alloys represent *metastable* states in supersaturated regions if and only if the precipitation is hindered kinetically, which also becomes difficult with increasing temperature. It is interesting to investigate whether truly equilibrium nanoalloys can exist in multicomponent alloys with HEGBs, which is suggested by Fig. 1a, but not yet verified experimentally.

Furthermore, a first-order premelting transition can result in a discontinuous increase in the GB excess entropy to accelerate the GB energy reduction with increasing temperature (Fig. 1a, albeit that studies showed GB structural disordering can increase GB mobilities.[41]). Likewise, a first-order adsorption transitions can also cause a discontinuous increase in the GB adsorption, which accelerates the GB energy reduction with increasing chemical potential (Fig. 2b).[23,26,44] It will be exciting to seek coupled GB disordering or adsorption transitions in HEGBs to further accelerate the $\gamma_{GB}$ reduction to potentially achieve zero GB energy within the solid solubility limit (to realize nanoalloys with equilibrium grain sizes at a true thermodynamic equilibrium).

We should develop quantitative models for the combined thermodynamic and kinetic stabilization of nanoalloys at high temperatures via utilizing HEGBs, following the concept shown in Fig. 3. We should also further explore the unique characters of HEGBs beyond thermodynamic and kinetic properties. A further extension is to investigate other types of high-entropy interfaces beyond GBs. For example, can we use high-entropy surfaces to stabilize nanoparticles for high-temperature catalysis or other applications?[47] Do high-entropy surfaces (or solid-solid heterointerfaces) have unique properties?

**Acknowledgement:** We gratefully acknowledge the support by the U.S. Army Research Office (Grant No. W911NF2210071, managed by Dr. Michael P. Bakas, in the Synthesis & Processing program).

**Author contributions:** J.L. conceived the initial concepts and basic theory of HEGBs and wrote the article. N.Z. assisted the detailed analysis to test the concepts and theory in his Ph.D. thesis with J.L..

**Competing interests:** The authors declare no competing interests.

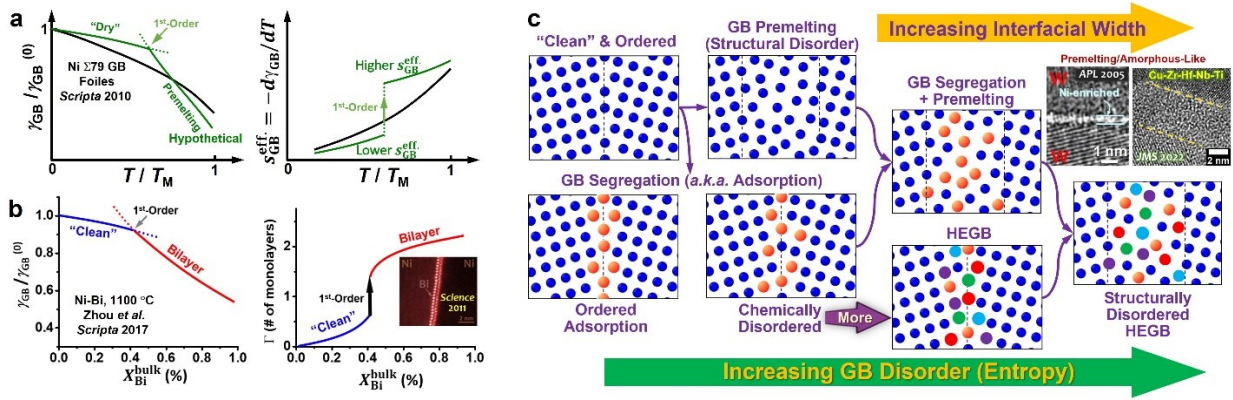

**Fig. 1. Grain boundary (GB) disorder, adsorption, and entropy. a** Normalized GB energy ($\gamma_{GB}/\gamma_{GB}^{(0)}$) and $d\gamma_{GB}/dT$ vs. normalized temperature ($T/T_M$) for a Ni Σ 79 GB (replotted after Foiles[21]) and a hypothetic case with a first-order GB premelting transition. **b** GB energy ($\gamma_{GB}$) and GB adsorption ($\Gamma$) vs. bulk Bi percentage ($X_{Bi}^{bulk}$) for a Bi-doped Ni general GB (replotted after Zhou et al.[23]), where the inset is a STEM image of Bi bilayer adsorption (reprinted after Luo et al.[26]). **c** Schematic illustration of different GB structures with increasing structural and/or chemical disorder and entropy. Insets are HRTEM images showing coupled GB premelting and segregation in Ni-doped W (reprinted after Luo et al.[27]) and a thicker amorphous-like GB complexion in Cu-Zr-Hf-Nb-Ti (reprinted after Grigorian & Rupert[20]).

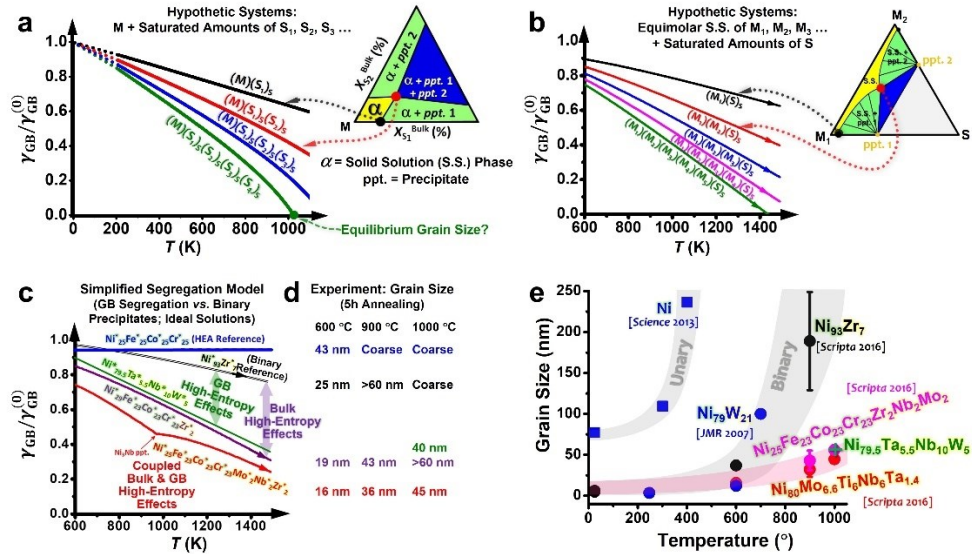

**Fig. 2. GB and bulk high-entropy effects. a** Computed normalized GB energy ($\gamma_{GB}/\gamma_{GB}^{(0)}$) vs. temperature ($T$) curves showing that a GB high-entropy effect can decrease $d\gamma_{GB}/dT$ along the solvus line in a hypothetic conventional multicomponent nanoalloy with one principal (M) and multiple segregating ($S_i$) elements. This GB high-entropy effect increases with increasing number of segregating elements. **b** Computed $\gamma_{GB}/\gamma_{GB}^{(0)}$ vs. $T$ curves along the bulk solvus curves, showing that $\gamma_{GB}$ can also be reduced via a bulk high-entropy effect that lower the bulk chemical potentials to suppress precipitation to promote GB adsorption for a hypothetic high-entropy nanoalloy of 1-5 principal ($M_i$) and one segregating element ($S$). In panel a and b, subscript "S" indicates the bulk composition of the segregating element is kept saturated (on the solvus line) at each temperature. **c** Computed $\gamma_{GB}/\gamma_{GB}^{(0)}$ vs. $T$ curves for several "Ni-like" alloys, where the superscripts "*" are used note that the calculations were conducted by using a lattice model of ideal solutions with segregation enthalpies and bonding energies to represent real alloys to capture important trends. **d** Selected experimental results of measured grain sizes of nanoalloys annealed at three different temperatures for 5 h, which shows a correlation between reduced GB energies and grain sizes. **e** Experimentally measured grain size vs. annealing temperature curves for several nanoalloys with HEGBs along with Ni and Ni-based binary nanoalloys for comparison.[9] The numerical and experimental results are replotted or regenerated using the same procedures, and the figure panels are reprinted or adapted, after Refs. 8,9, with permissions from Elsevier.

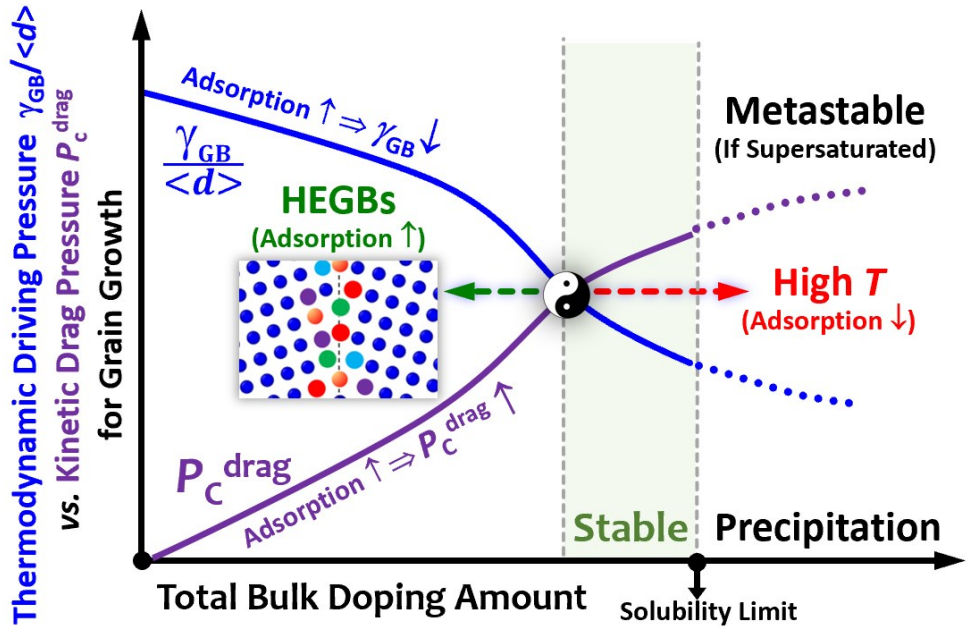

**Fig. 3. Stabilizing nanoalloys at high temperatures with high-entropy grain boundaries (HEGBs).** Schematic illustration of a balance of thermodynamic driving pressure ($\gamma_{GB}/\langle d \rangle$) and increase the critical kinetic solute drag pressure ($P_C^{\text{drag}}$) for grain growth, which result in the stabilization of a nanoalloy against grain growth. This balance has to be achieved before precipitation (below the solubility limit if the nanoalloy is at a thermodynamic equilibrium). Increasing temperature (*T*) can induce desorption (de-segregation) that moves the balance point towards right (to de-stabilize the nanoalloy, albeit the solubility limits also increase with increasing temperature). The HEGB-stabilized GB segregation can counter the thermally induced desorption. The increased total adsorption (the total amount of all segregating components) at HEGBs, which can be more stable at high temperatures with increasing number of components, can simultaneously reduce the thermodynamic driving force and increase the kinetic solute drag, thereby increasing the high-temperature stability (or moving the balance point towards left).